\documentclass[aps,floats,twocolumn,epsfig,nofootinbib,superscriptaddress,preprintnumbers]{revtex4}
\usepackage{amsfonts}
\usepackage{mathrsfs}
\usepackage{amssymb}
\usepackage{color}
\usepackage{epsfig}
\usepackage{graphicx}
\usepackage{amsmath}

 \newcommand{ \slashchar }[1]{\setbox0=\hbox{$#1$}   
    \dimen0=\wd0                                     
    \setbox1=\hbox{/} \dimen1=\wd1                   
    \ifdim\dimen0>\dimen1                            
       \rlap{\hbox to \dimen0{\hfil/\hfil}}          
       #1                                            
    \else                                            
       \rlap{\hbox to \dimen1{\hfil$#1$\hfil}}       
       /                                             
    \fi}                                             %
\newcommand{\met}{\slashchar{E}_T}

\newcommand{\lsim}{\lesssim}

\newcommand{\lstau}{\tilde\tau_1}

\newcommand{\sneut}{\tilde \nu_\tau}
\begin{document}

\vspace{10mm}
\title{\boldmath {$R$ parity violating SUSY explanation for the CDF $Wjj$ 
excess }}
\author{Dilip Kumar Ghosh}
\email{tpdkg@iacs.res.in}
\affiliation{Department of Theoretical Physics, Indian Association for the 
             Cultivation of Science, Kolkata 700032, India}
\author{Manas Maity}
\email{manas.maity@cern.ch}
\affiliation{Department of Physics, Visva-Bharati, Santiniketan 731235, India}
\author{Sourov Roy}
\email{tpsr@iacs.res.in}
\affiliation{Department of Theoretical Physics, Indian Association for 
             the Cultivation of Science, Kolkata 700032, India}
\begin{abstract}
Recently the CDF Collaboration has reported a statistically significant 
excess in the 
distribution of the dijet invariant mass between $120-160$ GeV in $Wjj$ 
event sample in $4.3~{\rm fb}^{-1}$ of data and later confirmed with 
$7.3 {\rm fb}^{-1}$ of data, which has generated considerable interest.
We offer a possible explanation of this observation in the general framework
of MSSM with $R$-parity violation through resonance production of 
$\tilde \nu_\tau$ decaying into the LSP $\tilde\tau_1$ and $W$ boson. 
We also give the predictions of this scenario for the LHC operating at 7 TeV 
center of mass energy.
\end{abstract}

\maketitle
The CDF Collaboration at the Fermilab has recently reported a $4.1\sigma $ 
excess in the dijet invariant mass distribution ($\rm 120 ~GeV <M_{jj}< 160 ~GeV$) in 
exclusive 2 jets + lepton + missing transverse energy ($\met$) events in $7.3{\rm fb}^{-1}$ 
of data from $p{\bar p} $ collisions at $1.96 ~{\rm TeV} $ where the 
lepton ($\ell = e, ~\mu $) and $\met$ are compatible with the decay of a real W \cite{CDFcollab}. 
A Gaussian fit to the excess region of $\rm M_{jj}$ indicates the peak to 
be around $\rm 147$ GeV and it is wider than the $Z$ boson of the Standard 
Model (SM). There are few hundred events in the excess region of 
$\rm M_{jj}$, which correspond to $\sigma (Wjj) \sim {\cal O}~(\rm pb)$. 
Such a large cross-section immediately rules out 
the possibility of the indication of the SM Higgs boson $H$, because 
$\sigma (p {\bar p} \to W^\pm H) \times {\rm BR}(H\to b {\bar b}) \simeq 12$ fb. 
The CDF analysis also reported no significant deviation from the SM 
expectation of $\rm M_{jj}$ distribution in $Z+$jets and 
$b\bar{b}\ell^{-}\met$ samples. This result is in agreement with the earlier 
published one \cite{Aaltonen:2011mk} by the CDF Collaboration with $4.3~{\rm fb}^{-1}$ of data 
sample, increasing the significance of the previous result. The position of the  
$\rm M_{jj}$ peak and the expected cross section for the process has not changed much.  

This anomalous excess in the $Wjj$ event sample has naturally generated a lot of 
interest in the particle physics community, as this could be the long-anticipated 
direct signal of new physics beyond the SM. However, the $\rm D0\!\!\!/$ Collaboration, 
using their $4.3~{\rm fb}^{-1}$ of data sample, claimed that the distribution of 
$\rm M_{jj}$ in such events is consistent with the SM prediction \cite {D0collab}.
Nevertheless, one should be rather cautious in either believing or disbelieving the claims 
made by the CDF or the $\rm D0\!\!\!/$ Collaborations. Before coming to any concrete 
conclusion, one must take into account the different methodologies used by the CDF and 
the $\rm D0\!\!\!/$ Collaborations to analyze their data samples, estimations of 
different systematics in the SM backgrounds, especially the QCD background. The 
$\rm D0\!\!\!/$ Collaboration by simulating $p {\bar p} \to WH \to \ell \nu b {\bar b}$ 
process to model acceptance and efficiency, put an upper limit of $1.9$ pb on 
the cross-section of anomalous dijet production at $95\% $ CL 
for $\rm M_{jj} = 145$ GeV \cite{D0collab}. As a result of this, their analysis 
does not rule out the possibility of new physics interpretation of the CDF 
dijet excess with a cross-section less than $1.9$ pb \cite{kribs}. 
On the other hand, one can argue that the CDF excess in dijet events 
could be due to some incorrect modeling of the QCD backgrounds in that mass window. 
To resolve this issue a joint task force has been formed \cite{jointtaskforce} and till 
we reach a definitive settlement of the issue it might be interesting to explore 
the different theoretical avenues that could account for the dijet anomaly. Many 
such possibilities have been already discussed 
recently \cite{kribs,Eichten:2011sh,kilic-thomas,varyinginterest,Sato:2011ui,Nelson:2011us}.

To explain the intriguing CDF result, we propose a scenario 
in $R$-parity violating (RPV) supersymmetric (SUSY) model, where the lightest stau 
(${\tilde \tau}_1$) is the lightest supersymmetric particle (LSP) and 
the next-to-lightest SUSY particle (NLSP) is the corresponding sneutrino, 
${\tilde \nu}_\tau$. All the other superpartners are rather heavy
$({\cal O}~(\rm TeV))$. The $\tilde\nu_\tau$ may be produced resonantly via 
$R$-parity violating coupling and subsequently decays into a real $\tilde\tau_1$ 
and an off-shell $W$ through $R$-parity conserving charged-current interaction. 
The real $\tilde\tau$ then decays into pair of jets via the same $R$-parity 
violating coupling and the off-shell $W$ decays into $\ell \nu$. Such resonant slepton 
production and decay has been briefly discussed in Ref.\cite{kilic-thomas} in the 
context of CDF $Wjj$ anomaly with a smaller mass splitting between the sneutrino and 
the stau (only due to the D-term splitting). As a result of this, the lepton from the 
decay of the virtual $W$ is rather soft to satisfy the CDF criteria of lepton 
selection which arises from the decay of real $W$. In another analysis 
\cite{Sato:2011ui}, authors assumed that the lepton 
and the neutrino come from the decay of the LSP charged slepton involving $R$-parity 
violating couplings and not from the W-boson. The sleptons are produced in the decay of 
the pair produced neutral winos. In addition they assumed that the wino and the charged 
slepton are nearly degenerate in mass and showed that this scenario can explain the 
CDF dijet anomaly. 

In our analysis of the proposed signal 
$p{\bar p} \to \tilde \nu_\tau \to W^- \tilde \tau_1^+ \to \ell^{-}\bar{\nu}_{\ell}jj$, 
we consider a large mass splitting between $\tilde \nu_\tau $ and $\tilde\tau_1$. 
This  may be achieved if one considers the ${\tilde \tau}_1$ to be a mixture of 
left-chiral stau (${\tilde \tau}_L$) and right-chiral stau (${\tilde \tau}_R$) 
and hence the mass splitting between the ${\tilde \nu}_\tau$ and ${\tilde \tau}_1$ can 
be made significantly larger ($\sim$ 200 GeV or so). To have $(\rm M_{jj})$ in the 
correct ball-park, we consider $\rm M_{\tilde \tau_1} $ in the range $140-150$ GeV. Now, 
the sneutrino (${\tilde \nu}_{\tau}$) produced through the $R$-parity 
violating coupling $\lambda^\prime_{311}$ decays into $\tilde \tau_1$ and an on-shell 
$W$, which then eventually lead to a pair of hard jets (from $\tilde \tau_1 $ decay) 
and a charged lepton and neutrino (from real $W$ decay) in the final state.  

In our numerical analysis, we use the CTEQ6L parton distribution function \cite{cteq6} 
with factorization scale $Q = \sqrt{\hat{s}}/2 $, where $\sqrt{\hat{s}}$ is the parton 
level CM energy. We also smear lepton and jet energies according to
\begin{equation}
\frac{\sigma (E_j)}{E_j} = \frac{a_j}{\sqrt{E/GeV}} \oplus b_j
\qquad
\frac{\sigma (E_\ell)}{E_\ell} = \frac{a_{\ell}}{\sqrt{E/GeV}} \oplus b_{\ell}
\label{resol}
\end{equation}
where, $a_{\ell} = 13.5\%, b_{\ell} = 2\%, a_{j} = 75\% $ and $b_{j} = 3\%$
\cite{Abazov:2007ev}.
After smearing we apply the selection criteria used by the CDF Collaboration,viz.
\begin{itemize}
\item 	lepton ($\rm\ell = e, ~\mu)$ and $\rm\met$ due to the neutrino should be consistent 
	with decay of a W boson: $\rm E_{T}^{e}(E_{T}^{\mu}) > 20 ~GeV$,  
	$\rm |\eta_{\ell}| < 1.0$ and $\rm\met > 25 ~GeV$ with transverse mass 
	$\rm \rm M_T(\ell \nu_\ell)  \rm > 30 ~GeV$.
\item	Two jets with $\rm E_{T}^{j} > 30 ~GeV$, cone size $\rm \Delta R = 0.4$;
	$\rm |\eta^{j}| < 2.4$ such that $\rm \mid \eta_{j_1} - \eta_{j_2}\mid  < 2.5$ 
	and $\rm (p_T)_{jj} > 40 ~GeV$.
\item	$ \mid \Delta \phi(\rm \met, j_1) \mid > 0.4 $ so that $\rm\met$ should not be due 
	to mismeasurement of jet energy   
\item 	lepton and the jets should be isolated, $\rm \Delta R(j,\ell) > 0.52 $.
\end{itemize}
\begin{figure}[htb]
\vspace{-10mm}
\begin{center}
\hspace{-8mm}
\includegraphics[scale=0.48]{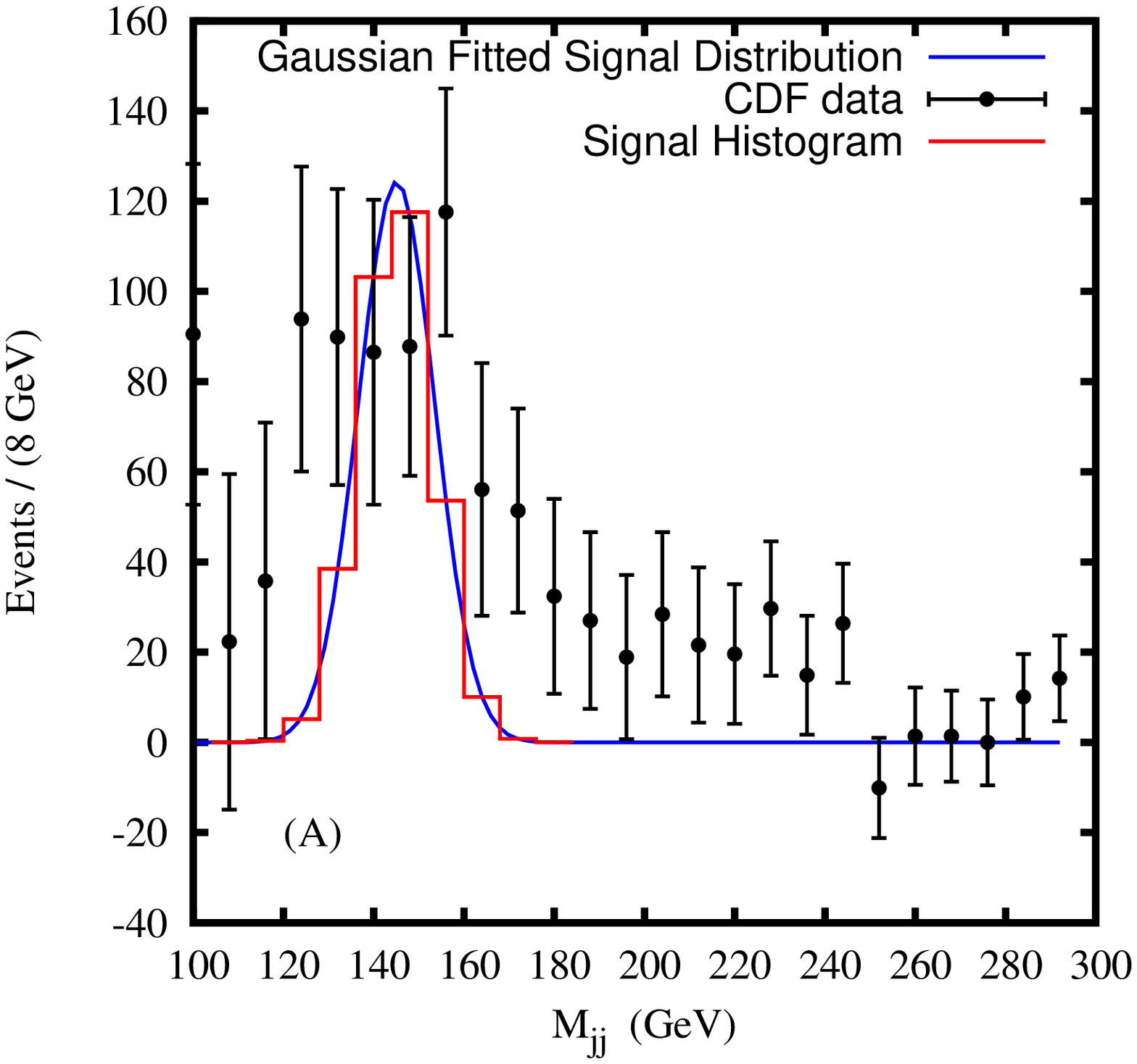}
\hspace{5mm}
\includegraphics[scale=0.48]{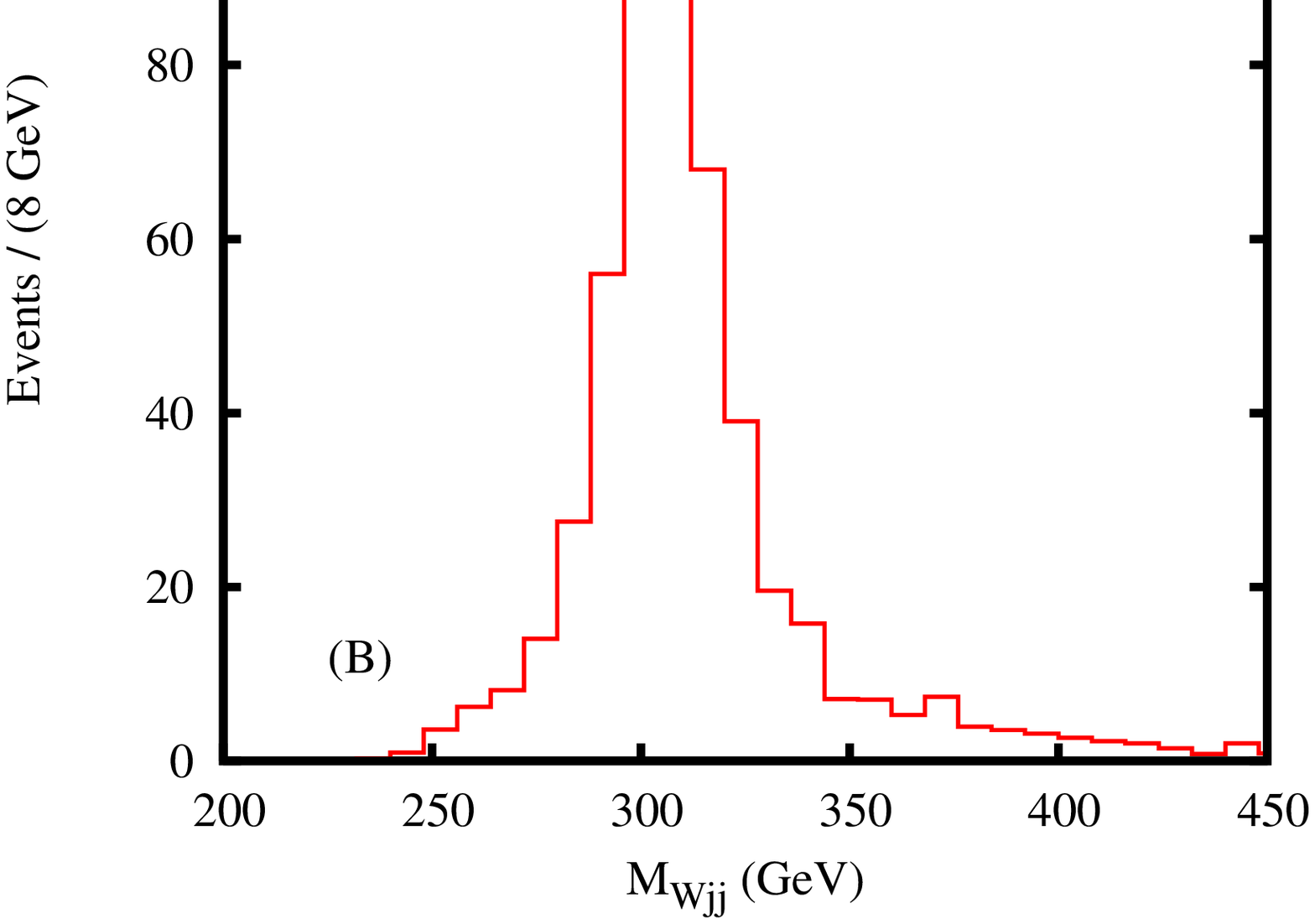}
\vspace{-70mm}
\caption{The top plot shows the distribution of dijet invariant mass $(\rm M_{jj})$ after the 
	CDF selection criteria have been applied: the points with error bars are taken from 
	CDF data (${\mathcal L}_{int} = 7.3 ~fb^{-1}$), the histogram (red) shows our 
	proposed signal events and a Gaussian is fitted to the histogram (blue line). 
	The bottom plot shows distribution of $\rm M_{Wjj}$ from our signal events.}
\label{jj_distribution}
\end{center}
\end{figure}

With all these, we are now well equipped to explain the observations of CDF. 
We set $\rm M_{{\tilde \nu}_\tau} = 300 $ GeV, $\rm M_{{\tilde \tau}_1} = 145 $ GeV. 
From this mass spectrum, one can see that $\tilde\nu_\tau$ can be produced 
in $s$ channel resonance via the $\lambda^\prime_{311}$ coupling\footnote{
Here, we assume only one $R$-parity violating coupling is dominant at a time.}
, which then decay into the LSP $\tilde \tau_1 $ and the gauge boson $W$. 
The LSP $\tilde\tau_1$ then decays into a pair of jets via the same $R$ parity 
violating coupling and the $W$ boson decays semileptonically, which eventually 
leads to $jj \ell \nu_\ell$ final state with low invariant mass ( $\rm M_{jj} \approx 140-150 $ GeV).


It has been shown in \cite{kilic-thomas} that the bound on $\lambda^\prime_{311}$ coupling from dijet 
resonance searches both at the CDF and UA2 experiments allows one to take a value 
of this coupling to be $\le$ 0.3. In our analysis we take the value to be 0.1. 
With these set of parameters, we estimate 
$\sigma_{\rm LO}  (p {\bar p} \to \tilde \nu_\tau\to W^- \tilde\tau_1^+\to jjW ) = 1.296$ pb,
which is  consistent with the $95\%$ CL upper limit on $\sigma (Wjj)$ by the
$\rm D0\!\!\!/$ Collaboration \cite{D0collab}. This signal cross-section (1.296 pb) 
eventually comes down to 71 fb after taking into account 
the leptonic ($e~\&~\mu)$ branching ratio of $W$ boson and imposing all the 
selection cuts mentioned above. There is no branching ratio suppression 
from $\tilde \tau_1 \to jj$ decay, as $\tilde\tau_1 $ decays into a pair of jets
with $100\%$ probability, which we think is a rather good approximation 
given the fact that all other SUSY particles are heavy. 

This cross-section corresponds to 518 (305) events with 
$\mathcal{L}_{int} = 7.3~{\rm fb}^{-1} (4.3~{\rm fb}^{-1})$. The $M_{jj}$ distribution according to our 
proposed scenario is shown in Fig. \ref{jj_distribution} (top plot). It is very 
clear that the signal histogram agrees reasonably well with the CDF data points. 
It may be noted that detailed implementation of detector effects is beyond the 
scope of this letter. Hence, instead of quantitative comparison of the number 
obtained in this analysis, which is somewhat larger than that of the CDF, we restrict 
ourselves to a qualitative comparison of the features of the distributions obtained. 
In the top plot of Fig.\ref{jj_distribution}, we normalize the signal histogram by the 
ratio of CDF peak value over the signal peak value. We also fit the signal histogram by 
a Gaussian and the mean of the Gaussian is given by $\rm M_{jj} = 145 $ GeV. The CDF Collaboration 
has given plots for several kinematic variables from their $7.3~{\rm fb}^{-1}$ data \cite{CDF_kinematic}. 
In the bottom plot of Fig.\ref{jj_distribution}, we display the invariant 
mass distribution of $\rm M_{\ell\nu jj}$ system. The longitudinal momentum $(p_z)$ 
of the neutrino is obtained by constraining $(p_\ell + p_\nu)^2 $ to $\rm M_W^2 $ 
and extracting the two possible solutions. To construct the $\rm M_{\ell \nu jj }$ we take the 
smallest solution between the two. The CDF Collaboration has shown a plot of $\rm M_{\ell \nu jj}$ 
after subtracting the SM background \cite{CDF_kinematic}. The plot shows 
excess in the region of our interest (250 GeV - 300 GeV). However, in their earlier analysis 
with $4.3~{\rm fb}^{-1}$ data sample, this $\rm M_{\ell \nu jj}$ distribution
was compatible in shape with the SM background only hypothesis \cite{Aaltonen:2011mk}.
As a result of this, one requires more statistics to reach a definite conclusion 
in $\rm M_{\ell \nu jj}$ distribution.

At this stage it is also important to look at some other interesting 
predictions of this scenario in the context of Tevatron. For example,
the pair production of $\lstau $ and subsequent decay of individual $\lstau $ via 
$\lambda^\prime_{311}$ coupling can lead to four-jet final state. We find that the leading 
order cross-section for $145$ GeV $\tilde \tau_1$ pair production is of the order 
of $(\sim 4-5~{\rm fb})$, which is rather small to be seen at the Tevatron with the
available statistics. The $\lstau$ can also be 
produced in association with $Z$ and Higgs bosons $H^\pm, h^0, H^0, A^0$ via 
$\lambda^\prime_{311}$ coupling. As before, $\lstau $ decays into a pair of jets via the 
same $R$-parity violating coupling, while $Z$ and Higgs bosons can have several
possible decay modes, which eventually lead to some spectacular signatures. Once again, 
for our scenario, the production cross-sections for these processes are too small to 
have any relevance at the Tevatron energy. For example, we find that the leading order 
cross-section $\sigma_{\rm LO}(p {\bar p} \to \lstau + Z) \sim 10^{-2} $ pb and
we expect that the cross-section for $\sigma (p {\bar p } \to \lstau \Phi^0),
~(\Phi = H^\pm, h^0, H^0,A^0) $ would be of the same order or even smaller 
(for heavier Higgs bosons) than $10^{-2}$ pb. In addition to these, one can have 
associated production of a ${\tilde \nu}_\tau$ and ${\tilde \tau}_1$ and this can lead 
to the $W + 4j$ signal where two pairs of jets will show peaks in their invariant 
mass distributions. Note that in this case the $2 \rightarrow 2$ production process 
is not suppressed by the small $R$-parity violating coupling. Similarly, pair-produced 
${\tilde \nu}_\tau$s can lead to $WW + 4j$ signal at the Tevatron. Note that,
with the mass spectrum considered here to explain the CDF dijet 
anomaly, one should not expect any significant number of events from
those two processes with the present luminosity at the Tevatron.

In conclusion in this letter we have analyzed the recently reported dijet invariant mass excess at $4.1\sigma $ 
in $M_{jj} \sim 120-160 $ GeV by the CDF Collaboration in $Wjj$ events with $7.3~{\rm fb}^{-1}$ data.
We have proposed an $R$-parity violating scenario with $\lambda^\prime_{311}$ as the dominant coupling, where 
$\tilde \nu_\tau$ and $\tilde\tau_1$ are the NLSP and the LSP respectively. In this scenario, the mass
splitting between the NLSP and the LSP is of the order of hundred GeV, such that NLSP can decay into
the LSP and a real $W$ boson. The resonant production of the NLSP can lead to a final state $\ell \nu jj $
where, the pair of jets coming from the $R$-parity violating decay of the LSP, show a peak at 
$M_{jj} \sim 145 $ GeV with a rather good agreement with the dijet invariant mass distribution
as shown by the CDF Collaboration at $7.3~{\rm fb}^{-1}$ data.
We have also found an $s$ channel resonance at $M_{\ell \nu jj} \sim 300 $ GeV,
corresponding to our chosen value of the NLSP mass. This result also appears to be very close
to the CDF plot of $M_{\ell \nu jj}$ after subtracting the SM background. However, for a 
definitive conclusion one requires higher luminosity. 
We note in passing that apart from 
the possible excess observed in $M_{\ell \nu jj}$ distribution around 250 GeV - 300 GeV, the
CDF excess in $Wjj$ events in $4.3~{\rm fb}^{-1}$ as well as $7.3~{\rm fb}^{-1}$ data samples
are almost consistent with each other. We have also discussed other possible signatures of this scenario at the 
Tevatron. 

Let us also comment briefly on the implications of this scenario at the LHC, which 
is accumulating $pp$ collision data at $\sqrt{s} = 7 ~{\rm TeV}$. We have found that the production 
cross-section for $Wjj$ from $pp \to \sneut \to \lstau + W $ is $7.3$ pb. At the LHC, 
the major SM background processes for this signal comes from the $W + n$ jets process 
($n \geq 2$). In the wake of the CDF result the ATLAS Collaboration performed a similar analysis 
with data from $pp$ collisions at $\sqrt{s} = 7 ~{\rm TeV}$ \cite{ATLAS_Note}. They have analyzed $33~{\rm pb}^{-1}$ 
of data collected in 2010 and did not observe any significant difference between SM 
expectation and observation in the mass range of interest. The estimated $W+n$ jets background
is approximately 20 times higher than the rate measured at the 
Tevatron \cite{ATLAS_Note}. After applying the same set of cuts as discussed before, 
the signal and the SM background cross-sections are $1.2$ pb and $34$~pb respectively. 
As a result of this, the signal significance of our scenario is $\lsim 1 \sigma$ at 
$33~{\rm pb}^{-1}$ of data. We know that both CMS and ATLAS have already 
collected $\mathcal{L}_{int} > 1 ~{\rm fb}^{-1}$ of data per experiment and are likely to 
collect a few ${\rm fb}^{-1}$ of data by 2012. Hence we may expect that with the increase of 
luminosity it might be possible to have a definitive conclusion on this issue from both 
CMS and ATLAS Collaborations.

\vspace*{1cm}
DKG thanks ICTP High Energy Group and the organizers of the Les Houches
workshop on ``Physics at the TeV Colliders'' for the hospitality where major
part of this work was done. MM acknowledges support from the Department 
of Science and Technology, India under the grant SR/MF/PS-03/2009-VB-I. 


\end{document}